\documentclass{article}

\usepackage{wrapfig}

\usepackage[nonatbib, final]{neurips_2024}

\usepackage[utf8]{inputenc} 
\usepackage[T1]{fontenc}    
\usepackage{hyperref}       
\usepackage{url}            
\usepackage{booktabs}       
\usepackage{amsfonts}       
\usepackage{nicefrac}       
\usepackage{microtype}      
\usepackage{xcolor}         

\usepackage[disable]{todonotes}
\newcommand{\chugo}[1]{\todo[color=red!60,inline]{H:#1}}

\title{Mind-to-Image: Projecting Visual Mental Imagination of the Brain from fMRI}

\author{Hugo Caselles-Dupr\'e$^{1}$, Charles Mellerio$^{2,3,4}$, Paul Hérent$^{5}$, Alizée Lopez-Persem$^{6}$,\\ \textbf{Benoit Béranger}$^{7}$, \textbf{Mathieu Soularue}$^{1}$, \textbf{Pierre Fautrel}$^{1}$, \textbf{Gauthier Vernier}$^{1}$, \textbf{Matthieu Cord}$^{8}$ \\
\textsuperscript{1}Obvious Research, Paris, France\\\textsuperscript{2}Neuroradiology Department, GHU Paris Psychiatrie et Neurosciences, Hôpital Sainte Anne \\ \textsuperscript{3}Université Paris Cité, Institute of Psychiatry and Neuroscience of Paris \\\textsuperscript{4}Centre Imagerie du Nord, Saint Denis, France\\\textsuperscript{5}Raidium\\\textsuperscript{6}FrontLab, Sorbonne Université, Institut du Cerveau\\\textsuperscript{7}Centre de NeuroImagerie de Recherche, Sorbonne Université, Institut du Cerveau\\\textsuperscript{8}Institut des Systèmes Intelligents et de la Robotique, Sorbonne Université \\
\texttt{\href{mailto:research.obvious@gmail.com}{research.obvious@gmail.com}}}

\AtBeginShipout
{\AtBeginShipoutAddToBox{
\tikz[remember picture,overlay] \node[opacity=0.4,inner sep=0pt] at (current page.center){\includegraphics[width=\paperwidth,height=\paperheight]{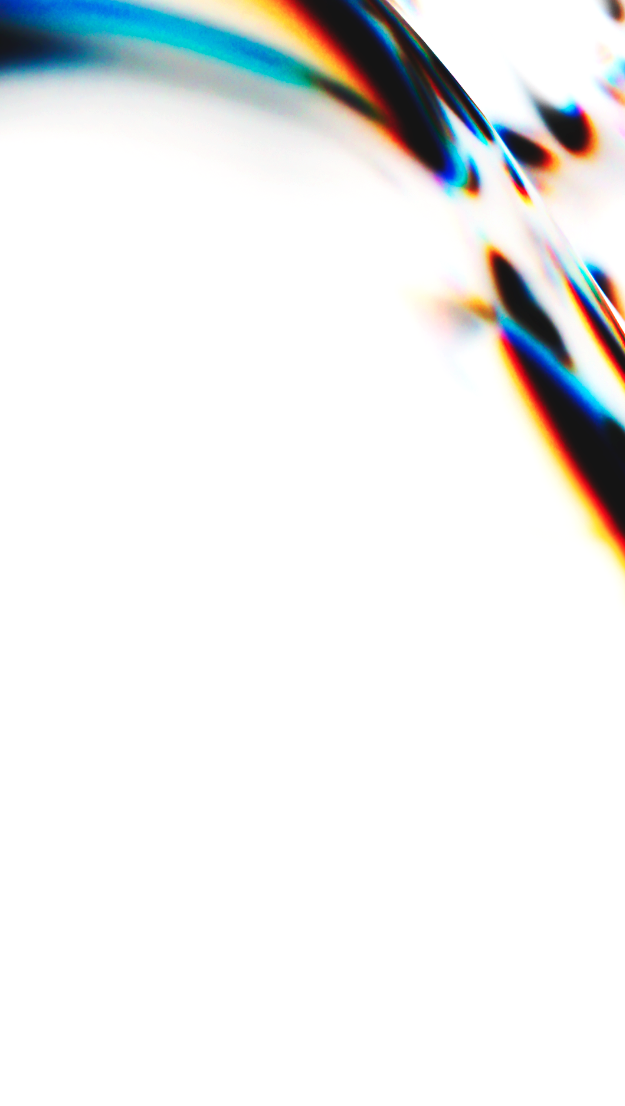}};
}
}

\begin{document}

\maketitle

\chugo{TODO list:Better figures and explanations for charles part, other approach: baseline with mindeye-v2 fine tuning, appendix: hyperparams}

\begin{abstract}

The reconstruction of images observed by subjects from fMRI data collected during visual stimuli has made strong progress in the past decade, thanks to the availability of extensive fMRI datasets and advancements in generative models for image generation. However, the application of visual reconstruction has remained limited. Reconstructing visual imagination presents a greater challenge, with potentially revolutionary applications ranging from aiding individuals with disabilities to verifying witness accounts in court. The primary hurdles in this field are the absence of data collection protocols for visual imagery and the lack of datasets on the subject. Traditionally, fMRI-to-image relies on data collected from subjects exposed to visual stimuli, which poses issues for generating visual imagery based on the difference of brain activity between visual stimulation and visual imagery. For the first time, we have compiled a substantial dataset (around 6h of scans) on visual imagery along with a proposed data collection protocol. We then train a modified version of an fMRI-to-image model and demonstrate the feasibility of reconstructing images from two modes of imagination: from memory and from pure imagination. The resulting pipeline we call Mind-to-Image marks a step towards creating a technology that allow direct reconstruction of visual imagery.

\end{abstract}

\section{Introduction}

\begin{figure*}[ht!]
    \centering
    \includegraphics[scale=0.2]{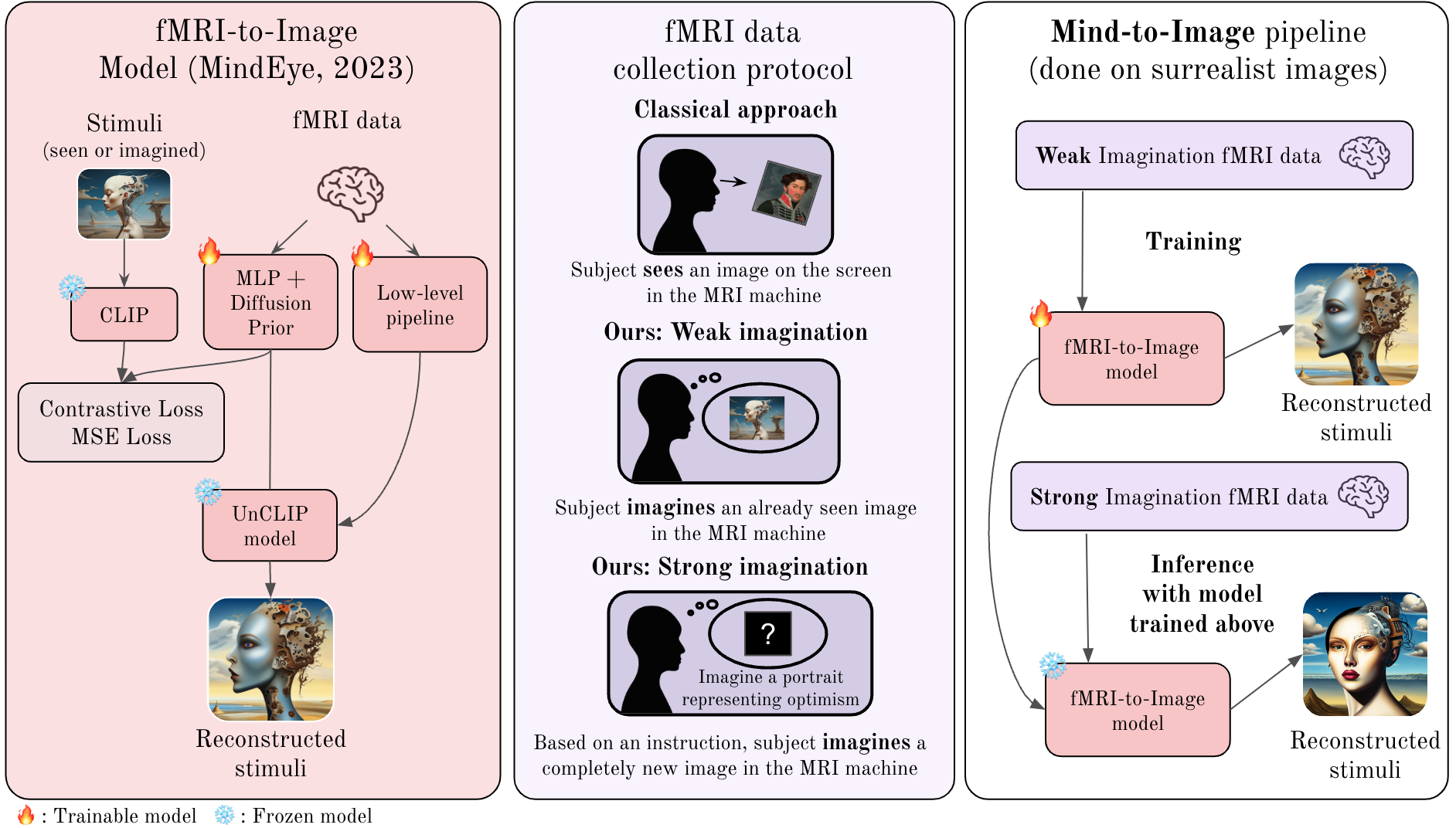}
    \caption{Overview of our Mind-to-Image pipeline which allows to reconstruct images from mental imagery. \textbf{Left}: overview of MindEye, the fMRI-to-Image model used in our Mind-to-Image approach. The stimuli (an image seen or in our case imagined) and the associated fMRI data is fed to a high-level pipeline (composed of an MLP and Diffusion Prior) and a low-level pipeline to create aligned CLIP embeddings of the fMRI data. \textbf{Middle}: fMRI data collection protocols. In the classic fMRI-to-Image approach, a subject sees images in an MRI scanner, BOLD (Blood-Oxygen Level Dependent) data are recorded then used to train an fMRI-to-Image model that converts brain data to match the seen images. In our approach, we devise two mental imagery protocols: weak and strong imagination. For weak imagination, we let the subject imagine previously seen images from a dataset of surrealist images (face portraits and nature landscapes) which we create. The brain data associated with the recollection of such images is gathered. For strong imagination, we collect brain data where the subject completely imagines new images based on instructions. \textbf{Right}: our Mind-to-Image pipeline. We use the weak imagination fMRI data to train an fMRI-to-Image model to reconstruct recollected images from visual imagination brain data. At inference time we use strong imagination fMRI data and this trained model to generate the imagined images, thanks to transfer learning.}
    \label{fig:main}
\end{figure*}

In the rapidly evolving field of neuroscience and artificial intelligence, the quest to decode the complex mechanisms of the human brain has led to groundbreaking developments, particularly in the area of functional magnetic resonance imaging (fMRI). Among the myriad applications of fMRI, one of the most intriguing has been the reconstruction of visual experiences. Traditional methods have successfully reconstructed images based on fMRI BOLD (blood-oxygen-level-dependent) data captured while subjects view specific images \cite{Chen2022SeeingBeyond, Chen2023CinematicMindscapes, Scotti2023ReconstructingMindsEye}. This process involves preprocessing the BOLD data to isolate voxels associated with the brain's visual areas, generating clip embeddings from these voxels, and then employing image generation algorithms that utilize these embeddings to recreate the observed images. Such methodologies have shown promising results, especially when employing datasets like the NSD (Natural Scenes Dataset) \cite{Allen2022MassiveFMRI}, a substantial 7T fMRI dataset designed to bridge cognitive neuroscience and artificial intelligence. This dataset, featuring extensive scans from eight subjects, has had tremendous impact for research in this domain.

Progress in cognitive neuroscience and artificial intelligence has not only allowed to advance our understanding of brain mechanisms but also to hold the promise to develop practical applications, such as real-time clinical tools and brain-computer interfaces. However, despite the progress in visual reconstruction from observed images, a challenge remains in the domain of visual imagination. Visual imagination encompasses both the recall of previously seen images (which we refer to as weak imagination in the paper) and the generation of entirely new images within the mind's eye (which we refer to as strong imagination). The ability to decode and reconstruct these imagined visuals offers a wide range of promising possibilities from aiding individuals with disabilities, to enhancing creative processes.

While using models based on reconstruction of images based on visual stimuli to generate images from imagination is theoretically possible, we believe that we need novel approaches to tackle the issue of visualizing imagination in the brain, particularly in data collection and model development. To this end, we introduce a protocol inspired by the NSD dataset but tailored to capture the nuances of visual imagination. Our approach distinguishes between weak and strong imagination, developing specific protocols for each type of visual imagery. We restrict our analysis on one theme of images: surrealism, and two distinct modalities of images: portraits and landscapes. This is done by collecting a dataset of surrealist face portraits and nature landscapes, using real and generated images. This choice is not only methodological but also thematic, aligning with the project's broader artistic aspirations. The intersection of art and science, particularly through the lens of surrealism and its exploration of automatic processes, provides a rich context for our research.

Leveraging the MindEye model \cite{Scotti2023ReconstructingMindsEye}, a state-of-the-art fMRI-to-image framework, we adapt its architecture to adapt to the larger dimensional complexity inherent in visual imagination data. Our approach is summarised in Fig.\ref{fig:main}. Our results, while preliminary, indicate the model's capability to, at least to some degree, generate portraits and landscapes based on the nature of the imagination. Although the fidelity of the reconstructions to the imagined content varies, which is measured quantitatively through usual metrics in the field, our findings underscore the potential of this approach. With further refinement and expansion of the dataset, this research could pave the way for more precise reconstructions of visual imagination, with the promise of a better understanding of the brain's creative and cognitive processes.

\section{Related work}

In the domain of mental imagery, several studies have sought to decode and reconstruct images from the brain's visual cortex when subjects engage in imagining specific scenes or objects. \cite{Stokes2009} provided early evidence of the brain's capacity to activate shape-specific population codes during mental imagery, highlighting the top-down influence of visual imagination on neural activity in the visual cortex. \cite{Reddy2010} explored the decoding of category information during mental imagery, providing an understanding on how different types of visual information are processed and represented in the brain.

More recently, \cite{Naselaris2015} developed a voxel-wise encoding model capable of decoding mental images of remembered scenes, demonstrating the relationship between memory, imagination, and visual perception. \cite{VanRullen2019} employs a combination of variational auto-encoders (VAE) and Generative Adversarial Networks (GAN) to perform pairwise decoding of faces with a high degree of accuracy but also to classify gender and decode which face was imagined by subjects. \cite{Goebel2022} showed the feasibility of reading imagined letter shapes from the mind's eye using real-time 7 Tesla fMRI, presenting a novel approach to understanding how specific visual information can be decoded from brain activity.

These studies collectively show the potential of fMRI data in decoding not just seen images but also those that are imagined or remembered.

More close to our work, a recent paper \cite{koide2024mental} provides preliminary results on reconstructing images from memory. While their results are encouraging, their reconstruction pipeline backbone is not state-of-the-art like MindEye is, which partly explains the results obtained. 

\section{fMRI data collection for visual imagery}
\label{sec:protocol}

The development of a comprehensive data collection protocol is one the main contributions of our study, aimed at capturing the complex neural correlates of visual imagination. Drawing inspiration from the NSD paper \cite{Allen2022MassiveFMRI}, our protocol is designed to explore two distinct modes of visual imagination: weak imagination (imagination from memory) and strong imagination (pure imagination). This approach enables us to study how the brain navigates between recalling visual information and generating novel visual constructs.

\subsection{Image dataset used: Surrealism}

We do not use the images from the COCO dataset \cite{lin2014microsoft} as it is done in the NSD dataset. Because our research is also associated with artistic research, we decided to narrow the dataset to one theme (surrealism) and two modes (face portraits and nature landscapes). This choice narrows the visual space of the fMRI-to-Image model, which simplifies the task and allows us to have results without having to do 40 hours of scans as in the NSD dataset.

Thus, we created our own dataset of images for our experiments. The images come from three sources: real images and generated images using Versatile Diffusion \cite{Xu2023VersatileDiffusion} and Midjourney. The dataset comprises 1200 images.

This dataset is used in the weak imagination protocol, where the images are recollected from memory and imagined. Then, the images are used to train the fMRI-to-Image model. Finally, only brain data is given to the fMRI-to-Image model to produce the reconstructions, both in the weak and strong imagination cases.

\subsection{fMRI parameters}

Data was collected on Siemens 3T PrismaFit using a 64 channels headcoil. BOLD data were acquired using the CMRR multiband EPI sequence. The main parameters are: 2mm isotropic voxels with full brain coverage, TR/TE = 1300ms/27ms, MultiBand=4, partialFourier=7/8, FlipAngle=68°, P$>>$A phase encoding direction, EchoSpacing at minimum.

\subsection{Weak imagination: data collection protocol}

The protocol for weak imagination aims at recording brain activity corresponding to the recall of previously seen images. In this phase, subjects are first shown an image for 3 seconds, followed by a 1-second rest period. This sequence is repeated with three different images, each selected from our curated set of 600 surrealist portraits and 600 surrealist landscapes. To engage the brain's imagination, the same images are then flashed for 0.1 seconds, prompting the subject to imagine the image for 5 seconds before resting for another second. This process is repeated over a span of 6 hours, encompassing a total of 1200 images. Out of the data collected, 75 data points (comprising brain activity and the corresponding image) are reserved for validation and evaluation, with the remainder serving as the training set for our fMRI-to-image model. The evaluation focuses on the model's ability to generate images that not only match the correct modality (face portrait or nature landscape) but also reflect the contents of the original image.

\subsection{Strong imagination: data collection protocol}

For strong imagination, the subject is now asked to generate entirely new visual content based on verbal prompts. The subject is instructed to imagine a portrait or landscape evoking one of ten primal emotions (such as fear, happiness, love, etc.). This task is designed to stimulate the pure imagination process, encouraging the brain to construct novel visual imagery without the reference of previously seen images. Each imagination session lasts for 6 seconds, followed by a 1-second rest, cycling through combinations of landscape/portrait and emotional prompts. This cycle is repeated 10 times. Unlike the data collected for weak imagination, this dataset is solely used during the model's inference stage. The evaluation here is similar to the weak imagination evaluation: quantitatively assess if the model matches the correct modality (face portrait or nature landscape) and qualitatively assess if the contents of the generated image matches the visual imagery of the subject. However we cannot compute all the quantitative metrics as the ground truth image does not actually exist outside the subject's mind. To ensure that we recall what the subject has imagined, the subject provides an oral description of the visual contents of the imagined image.

\subsection{Pre-processing of raw BOLD data and masking of the regions of interest}

After the raw  BOLD (blood-oxygen-level-dependent) data is collected, we apply a preprocessing routine to provide the model with data that decorrelates all the noise that impair the following steps. We first co-register all BOLD data with a T1 anatomical scan. Then, using a General Linear Model (GLM) with the Glover hemodynamic response function (HRF) (implemented via the Nilearn library \cite{Abraham2014MachineLearningNeuroimaging}), we extracted beta values associated with each voxel for every distinct event, including both the displayed images and the periods of imagination. This allows us to isolate the specific brain activity corresponding to each event, which is common practice for such tasks.

\textbf{Mask selection.} In the subsequent data pre-processing phase, we identify the brain regions most important to visual imagination and visual perception, using a GLM model to manually extract these areas of interest. Unlike the NSD dataset, which predominantly focuses on visual areas, our study required the creation of a custom mask to englobe the regions identified as crucial in our context of weak and strong imagination. 

It is a well known fact that the generation of the perceptive or imagined visual experience \cite{dijkstra2019shared} involves the ventral stream cortical areas (in occipital and temporal lobe). In these visual regions, various aspects of visual information contribute to creating a mental image. This includes features like shape, color, motion, depth, and texture. These features are processed and integrated across low-level (V1,V2), associative (V3, V4, V5) and high-level (including fusiform gyrus in temporal lobe) cortical areas to form a coherent mental representation of what we see or what we think of. 

This is the reason why we designed the brain mask to collect the cortical responses during the imagination conditions using the mean responses of the general linear model applied to the event vs. rest condition, in the occipital and temporal lobe. This method aimed to consider only the cortical activity involved in creating a mental image in one subject during the imagination phase.

We thus create and compare three masks manually curated based on the GLM analysis results, ensuring that only voxels corresponding to our defined regions of interest were included in further analyses. The delineation and analysis of these regions, and how they diverge from the predominantly visual areas highlighted in the NSD study, is studied in the results section.

Finally, we create our training, validation (weak imagination) and test (strong imagination) datasets by flattening the pre-processed betas corresponding to the selected voxels and pair them with the images or prompts related to their events.

\section{Mind-to-Image pipeline}

We now describe the fMRI-to-Image model used for our experiments and then outline our Mind-to-Image pipeline by connecting the dots between the data collection protocol and the fMRI-to-Image model.

\subsection{fMRI-to-Image model}

We experiment with two approaches for the choice of the fMRI-to-Image model using in our Mind-to-Image pipeline.

\subsubsection{Specialized model approach} 

We start from the state-of-the-art MindEye model which reconstructs images viewed by subjects from fMRI data by combining two pipelines: a high-level (semantic) pipeline and a low-level (perceptual) pipeline. The high-level pipeline maps fMRI voxels to the CLIP Vision Transformer L/14 image space, capturing the semantics of the image. The low-level pipeline targets the perceptual fidelity of reconstructed images by mapping voxels to Stable Diffusion's Variational Autoencoder (VAE) embedding space. They leverage contrastive and mean squared error (MSE) losses for training of its projection and reconstruction capabilities. At inference time, fMRI data associated to an image is fed through the model which outputs a low-level reconstructed image and aligned CLIP embeddings. These embeddings are fed to an UnCLIP model, which is a model able to reconstruct an image based on its CLIP embeddings. In this case, the embeddings as well as the low-level reconstructed image is fed to a Versatile Diffusion Image Variations pipeline to obtain the final reconstruction, as it is done in the original MindEye paper. The whole fMRI-to-Image pipeline is illustrated in Fig.\ref{fig:main}.

\quad \textbf{Adapting fMRI-to-Image to larger masks for visual imagery.} To accommodate the increased dimensionality inherent in our fMRI BOLD voxel data, which covers extensive brain areas notably larger than only the visual areas of the brain, we adapted MindEye's architecture. This adjustment aimed to manage the growth in input dimensions without exponentially increasing the model's parameter count. We thus simply refine the architecture of the MLP so that the model efficiently processes the data while maintaining a reasonable cost in GPU memory.

\subsubsection{Alternative approach: fine-tuning from pre-trained fMRI-to-Image model on vision.} 

Another approach we experimented with is to leverage pre-trained multi-subject models on a vision task on the NSD dataset. Then, we hope that fine tuning these models with our imagination data will help the model generalize better, motivated by the overlap between the vision and imagination tasks. We thus experiment with MindEye2 \cite{scotti2024mindeye2}, the recent update to MindEye which among several technical improvements provides a multi-subject model on the NSD dataset (vision task). This model can be fine-tuned given user data, which we proceed to do in our experiments. However there are a few issues to consider: the data is not collected the same way as the NSD data, nor pre-processed exactly the same, which we posit can hinder the approach. 

\subsection{Mind-to-Image pipeline} 

Our Mind-to-Image experimental procedures are as outlined below:

\begin{itemize}
    \item We first collect a dataset of weak and strong imagination using the protocol defined in Sec.\ref{sec:protocol}, with a total of 6 hours of scans.
    \item Then, we train our adapted fMRI-to-Image model using the dataset obtained from the weak imagination data collection protocol. Then, this model is evaluated using the validation set to assess its performance.
    \item We finally freeze the parameters of the fMRI-to-Image model. This frozen model is then employed to perform inference on the dataset derived from the strong imagination data collection protocol, with the goal of visualising the purely imagined images thanks to transfer learning.
\end{itemize}

\section{Results}

We first analyze the most activated regions of the brain during visual perception, weak imagination and strong imagination. 

We then go over the results obtained for training reconstruction fMRI-to-Image models for weak imagination before looking at strong imagination.

\subsection{What happens in the brain during visualisation, weak and strong imagination?}

\begin{wrapfigure}{l}{0.5\textwidth}
    \begin{center}
    \includegraphics[scale=0.23]{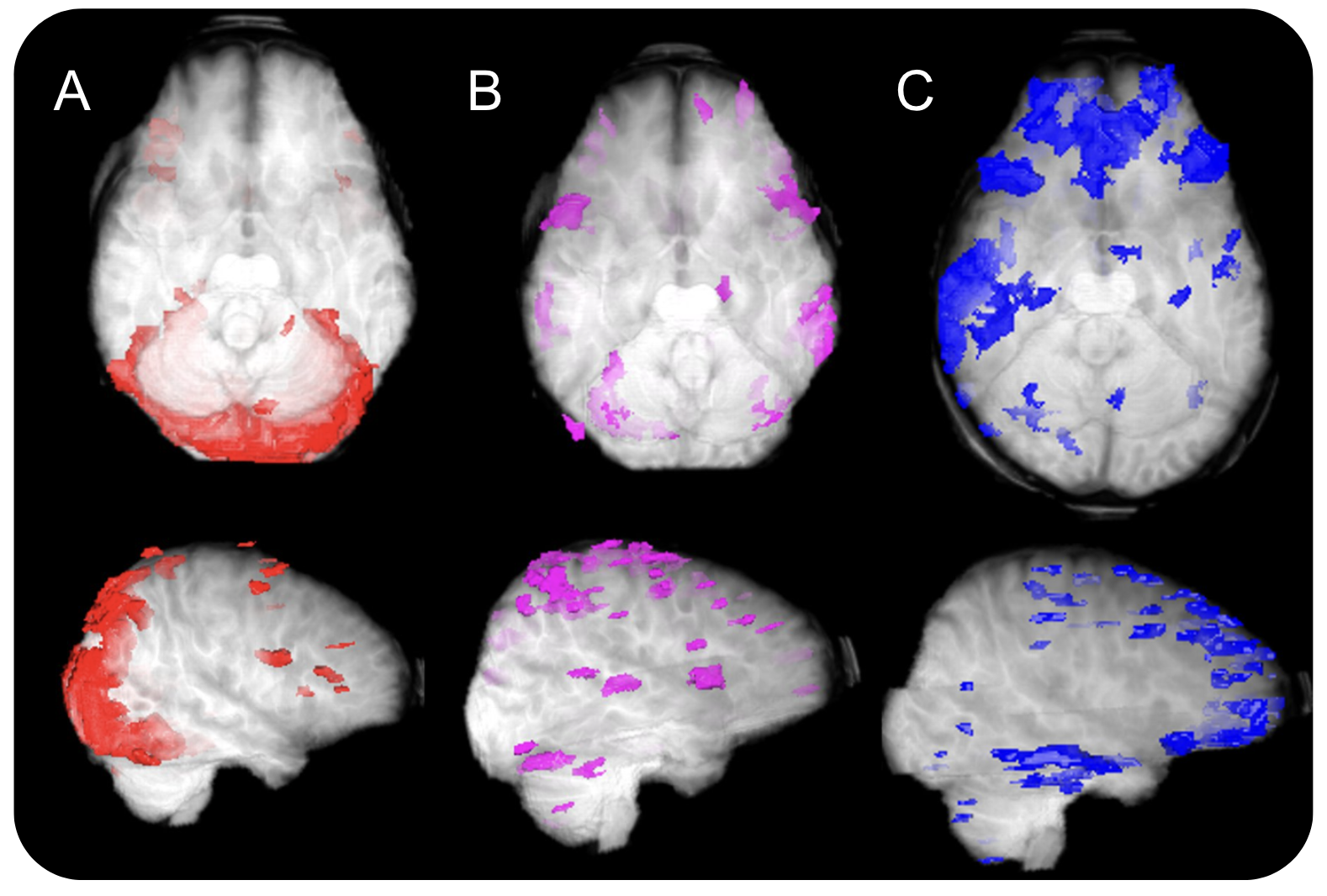}
    \end{center}
    \caption{Functional MRI results of the three conditions, respectively: simple visual perception (A- in red), weak imagination (B- in pink) and strong imagination (C- in blue) presented through a lower (top row) and right lateral (bottom row) view. Activation maps were generated by comparing the “test” and “rest” conditions, using the same statistical threshold for each test (T-score =5).}
    \label{fig:brains}
\end{wrapfigure}

In this section we describe the brain regions involved during visual perception, weak imagination, and strong imagination. This is done through the visualisations presented in Fig.\ref{fig:brains}.

 The results of the first test (A) show that the brain regions 
involved in normal visual perception are, as expected, the primary and secondary visual areas in the occipital lobe, the superior parietal lobule and the bottom of the temporal lobe. 


Weak imagination (B) involves occipital primary visual areas in a much weaker way. Instead, there is an activation of the basal temporal associative visual areas (fusiform gyrus) involved in mental imagery. The latter are partially overlapped with brain areas involved in visual perception. 

Strong imagination (C) involves even more brain areas distinct from visual cortex, with a strong activation in prefrontal regions, involved in the strategy of generating and controlling mental imagery. There is also an incomplete overlap of activations between visual perception and imagination in the temporal lobe, mainly limited to the fusiform gyrus, especially in the right hemisphere, more specialized in the face processing (as this result was obtained during the portrait imagination test).

These results justify the need for a specialised dataset for imagination relative to visual perception, given that the regions most active during the different tests are not quite the same.

We thus devise three masks that will be used to select the voxels of the brain that are given as input to the fMRI-to-Image model. \textbf{Mask Vision} (based on the activations during visual stimuli), \textbf{Mask Imagination} (based on the activations during weak imagination), and \textbf{Mask Vision$\cup$Imagination} (union of A and B, englobing areas associated to visual perception and weak imagination).

\subsection{Can we reconstruct images from weak imagination / memory?}

\begin{figure}
\centering
\begin{minipage}{.45\textwidth}
    \centering
    \includegraphics[scale=0.17]{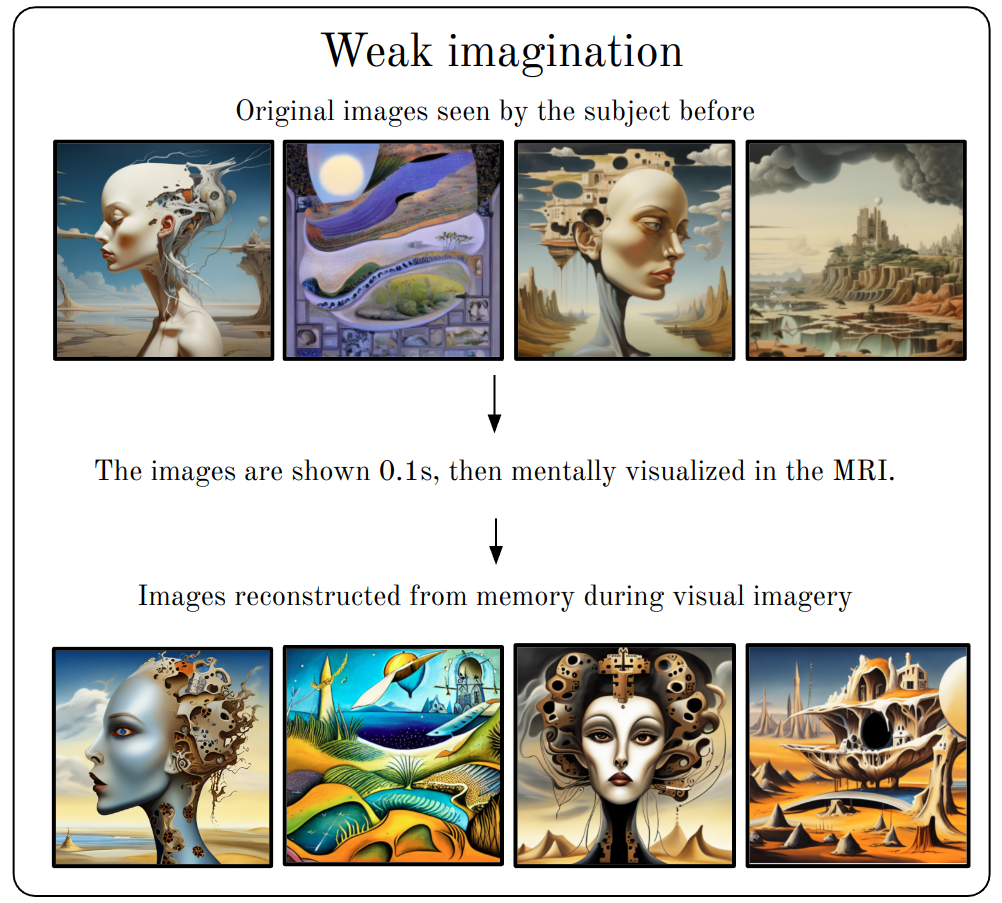}
    \caption{Results for weak imagination on the validation dataset. Top row: images from the validation set of our surrealism image dataset, which were recollected from memory by the subject in the fMRI scanner. Bottom row: reconstructed images from the fMRI-to-Image model based on the brain data associated with the recollection of the images in the top row.}
    \label{fig:weak}
\end{minipage}%
\hspace{1cm}
\begin{minipage}{.45\textwidth}
    \centering
    \includegraphics[scale=0.15]{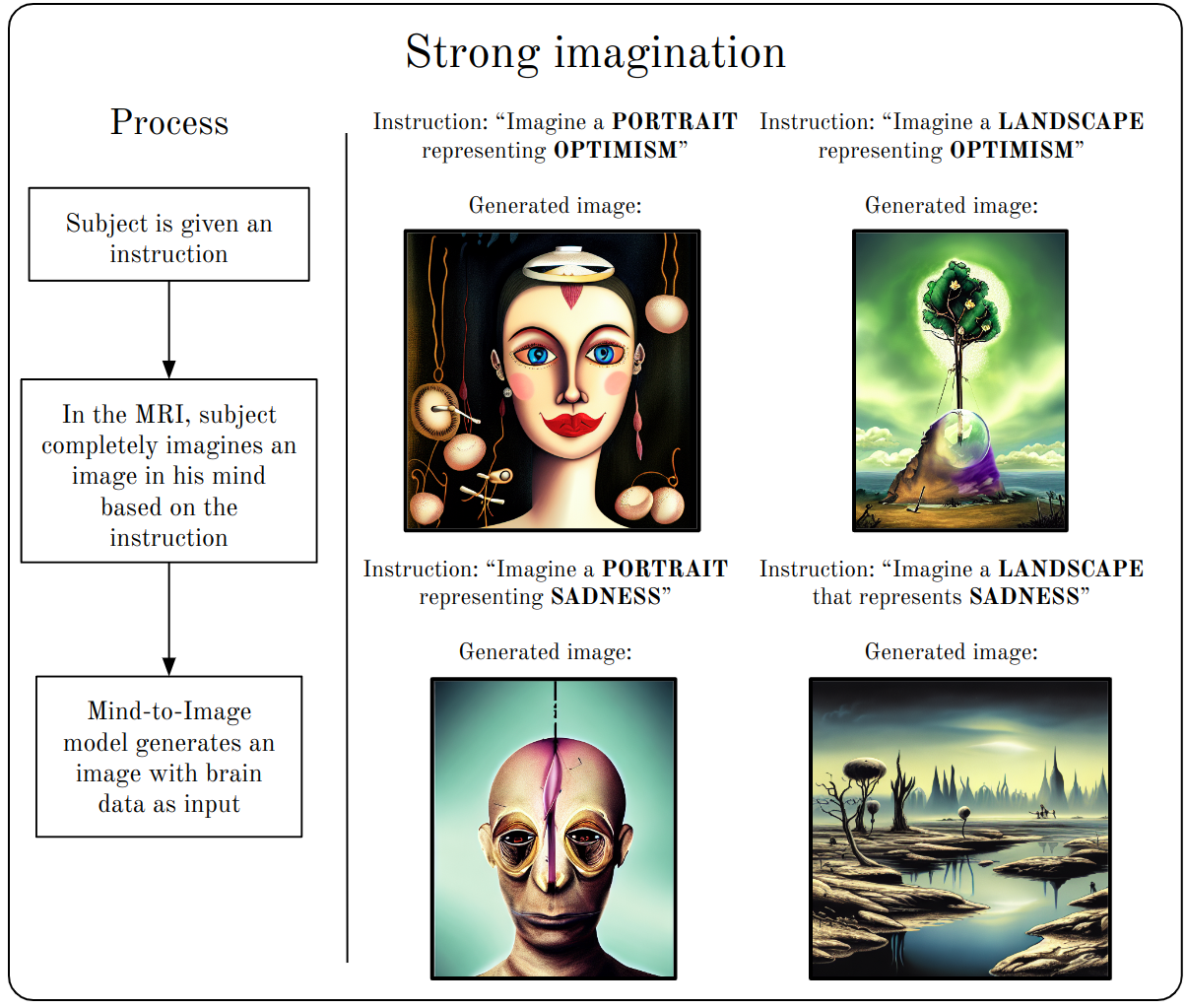}
    \caption{Results for strong imagination. Subject is given an instruction on the fMRI screen, such as "Imagine a portrait representing optimism". Then, the subject purely imagines such an image, giving an oral description. Then, the associated brain data is fed to the fMRI-to-Image model to produce a reconstruction, evaluated compared to the oral description.}
    \label{fig:strong}
\end{minipage}
\end{figure}

We present curated results of the images generated by our model from weak imagination in Fig.\ref{fig:weak}.

\quad \textbf{Modality assessment.} In the exploration of weak imagination, where the subject is asked to recall previously seen images, our first observation is that the model is able to identify the class of the images. Specifically, if a subject recalled a face portrait, the reconstructed image is almost always a face portrait, and similarly for nature landscapes. This observation suggests the ability of the model to grasp the category of the imagined content accurately within the BOLD data. We confirm this quantitatively by fine-tuning a pretrained ResNet50 model \cite{he2016deep} to classify images as either face portraits or nature landscapes using the images in our dataset. We apply this classifier to the reconstructed images from the weak imagination tasks of the validation set. We obtain, on our best performing model (MindEye1 - Mask Vision$\cup$Imagination), a classification accuracy of 91\% in correctly identifying the reconstructed images as belonging to their original categories of either portraits or landscapes. This first result demonstrates the model's ability to identify the modality in the brain data during weak imagination.

\begin{table*}[ht!]
    \centering
    \setlength{\tabcolsep}{1pt}
    \small
    \begin{tabular}{lcccccccc}
        \toprule
        Stimuli \& Dataset & \multicolumn{4}{c}{Low-Level} & \multicolumn{4}{c}{High-Level} \\
        \cmidrule(lr){2-5} \cmidrule(l){6-9}
         Model & PixCorr $\uparrow$ & SSIM $\uparrow$ & Alex(2) $\uparrow$ & Alex(5) $\uparrow$ & Incep $\uparrow$ & CLIP $\uparrow$  & Eff $\downarrow$ & SwAV $\downarrow$ \\
        \midrule
        \midrule
        Vision results on NSD dataset &&&&&&&& \\
        \cmidrule(lr){0-0}
        MindEye \cite{Scotti2023ReconstructingMindsEye} & ${.309}$ & ${.323}$ & ${94.7\%}$ & ${97.8\%}$  & ${93.8\%}$ & ${94.1\%}$ & ${.645}$ & ${.367}$ \\
        \midrule
        \midrule
        Mind-to-Image &&&&&&&& \\
        (Weak imagination on Surrealism dataset) &&&&&&&& \\
        \cmidrule(lr){0-0}
        MindEye1 - Mask Vision & ${.177}$ & ${.051}$ & ${57.8\%}$ & ${53.0\%}$  & ${58.4\%}$ & ${57.9\%}$ & ${.718}$ & ${.079}$ \\
        MindEye1 - Mask Imagination & ${.082}$ & ${.023}$ & ${54.8\%}$ & ${51.6\%}$  & ${55.1\%}$ & ${54.9\%}$ & ${.797}$ & ${.098}$ \\
        MindEye1 - Mask Vision$\cup$Imagination & ${.165}$ & ${.052}$ & ${\mathbf{65.1\%}}$ & ${\mathbf{66.4\%}}$  & ${\mathbf{61.4\%}}$ & ${\mathbf{68.5\%}}$ & ${.764}$ & ${.065}$ \\
        MindEye2 Multi-Subject Fine-Tuning &&&&&&&& \\ - Mask Vision$\cup$Imagination & ${.179}$ & ${.064}$ & ${49.1\%}$ & ${48.4\%}$  & ${51.2\%}$ & ${47.8\%}$ & ${.661}$ & ${.097}$ \\
        \bottomrule
    \end{tabular}
    \caption{Quantitative comparison of Mind-to-Image reconstruction performance against other models trained on different datasets, computed on usual metrics in the field. The results are not directly comparable, but are given to provide an element of comparison on the ranges of the values of the metrics. PixCorr is the pixel-level correlation of reconstructed and groundtruth images. SSIM \cite{wang2004image} is the structural similarity index metric. AlexNet(2) and AlexNet(5) are the 2-way comparisons of the second and fifth layers of AlexNet \cite{krizhevsky2012imagenet}, respectively. Inception is the 2-way comparison of the last pooling layer of InceptionV3 \cite{szegedy2016rethinking}. CLIP is the 2-way comparison of the output layer of the CLIP-Vision model \cite{radford2021learning}. EffNet-B and SwAV are distance metrics gathered from EfficientNet-B \cite{tan2019efficientnet} and SwAV-ResNet \cite{caron2020unsupervised} models, respectively.}
    \label{tab:recon_eval}
\end{table*}

\quad \textbf{Quantitative results.} We provide in Tab.\ref{tab:recon_eval} the usual metrics to quantitatively assess the performance of reconstruction from weak imagination. As an indication, we compare our results with MindEye \cite{Scotti2023ReconstructingMindsEye}. 
Note that the comparisons are not apples-to-apples: \cite{Scotti2023ReconstructingMindsEye} provide results for visual stimuli experiments (not for weak imagination) on a large-scale (approx. 10x size) dataset that uses another image domain (realistic images), explaining the big gap in performance. Our results, while inferior to those of reconstructions from visual stimuli, show that we are able to have significantly better than chance results on reconstruction from weak imagination. 

The best results are obtained when using our specialized model approach with MindEye1 in combination with \textbf{Mask Vision$\cup$Imagination}. This shows the usefulness of considering both the visual areas and weak imagination areas of the brain. Specifically, our results indicate that the importance of visual area is superior relative to the imagination areas, as the model trained with \textbf{Mask Vision} outperforms the one trained with \textbf{Mask Imagination}. The approach of fine-tuning from a multi-subject model trained on a vision tasks is not conclusive, as the obtained results are not better than chance. For now, it is not possible to conclude on the relevance of such approach, because the poor performances can be due to data distribution differences and data pre-processing procedures. 

Regarding the metric scores, correlations/distance between the reconstructed and original images (PixCorr, SSIM, Eff and SwAV) do not seem representative of model performance compared to 2-way comparisons (Alex(2), Alex(5), Incep and CLIP). For instance, MindEye2 multi-subject fine-tuning scores extremely well on correlations/distance metrics, while qualitatively we observed poor results. MindEye2 uses SD-XL for its image generation, which is stronger than Versatile Diffusion used in MindEye1: this could explain the result as the generated images with MindEye2 are naturally better looking. For clarity purposes, we decided not to bold the results for correlation/distance metrics and based our comparisons on the 2-way comparisons metrics. The quality of metrics is a known problem in the field and some authors have shifted toward human-based assessments, which can be hard to reproduce under the same conditions. Further investigation remains needed to assess the quality of the metrics used in the field.


\quad \textbf{Qualitative results.} While the model demonstrated a robust ability to regenerate the broad category of the original images, the fidelity of content reconstruction within these images presented a more varied outcome. Some reconstructions were visually semantically close to the remembered images. However, others diverged more significantly, indicating a variability in the model's ability to precisely capture and reconstruct the nuanced details of each imagined scene. We provide uncurated reconstructions in Fig.\ref{fig:uncurated-weak} in Appendix. This variance in reconstruction fidelity can be attributed, in part, to the limitations imposed by the dataset's size and the duration of the fMRI scanning sessions. With only 6 hours of scans and 1200 images constituting the dataset, the goal for training a model from scratch to achieve perfect reconstruction is out of reach. Nevertheless, these initial results are promising, suggesting a capacity for accurate content reproduction. It is reasonable to anticipate that with an expanded dataset, the fidelity of the reconstructed images to their original counterparts would see important improvement, enhancing both the precision and consistency of the content reproduction in future iterations.

\subsection{Can we reconstruct images from strong imagination / pure imagination?}

We present curated results of the images generated by our model from strong imagination in Fig.\ref{fig:strong}.

In the phase dedicated to strong imagination, we employed the previously trained and then frozen fMRI-to-Image model to process BOLD data obtained while subjects engaged in pure imagination tasks, such as imagining a portrait or landscape related to a broad concept like optimism. We thus rely on transfer learning, leveraging a model trained exclusively on weak imagination data to interpret and generate visual outputs from the distinct context of strong imagination, without fine-tuning. Indeed, we do not have tangible images to serve as direct training data for strong imagination, since these images exist solely within the mind of the subject. Hence we employ transfer learning from the weak imagination process to generate from strong imagination. Remarkably, the application of the fine-tuned ResNet50 classifier to the generated images resulted in an 88\% accuracy rate in categorizing the images according to the instructions provided to the subjects. This accuracy in category generation demonstrates the effectiveness of transfer learning from weak to strong imagination, which represent our main contribution.

Determining the content accuracy of images generated from strong imagination poses a unique challenge, as the 'ground truth' images solely exist in the subject's mental imagery, inaccessible for direct comparison. To navigate this, we based our evaluation on subjects' descriptions of their imagined scenes, collected during or following the scanning sessions. The variability in the reconstructed content ranges from highly accurate renditions of specific imagined elements, like a tree, dark-haired woman, or flowers, to more abstract interpretations or plain errors. This also reflects the subjective nature of this task. Given the reliance on subjective descriptions and the inherent open-endedness of strong imagination, the results should be interpreted with caution. While it's challenging to draw definitive conclusions about the model's capacity for precise content reconstruction in this context, the instances of recognisable elements emerging in the reconstructions indicate what could be achievable with increased data volume and further model refinement.




\section{Discussion}

\subsection{Ethical considerations}

As we explore further into the capabilities of decoding and reconstructing visual content from brain activity, we get closer to the boundaries of personal thought and imagination, a concept known as "mind privacy". The ability to visualise an individual's mental imagery raises questions about consent, mental autonomy, and the potential for misuse of such technologies. On the one hand, such technology provides potential for good, opening avenues for novel communication methods for those unable to speak or write, enhancing creative expression, and deepening our understanding of the human brain and its processes. On the other hand, it introduces the possibility of invasive breaches of personal thought, where the most private mental images and ideas could be exposed without consent. This requires a careful, deliberate approach to ensure that the development and application of these technologies are governed by ethical principles that prioritise individual privacy and autonomy. Currently, the barriers to entry, including the need for explicit consent and the use of expensive and sophisticated MRI machinery, provide a layer of protection.

\subsection{Future work}

Future directions for this line of research include the acquisition of more extensive data to improve the accuracy and specificity of content match in the reconstructed images. A refined experimental protocol for both weak and strong imagination tasks could lead to insights into the brain's imaging processes, while improved evaluation methods for strong imagination would enhance our understanding of the model's effectiveness in capturing the depth and detail of imagined content. 

Additionally, the question of whether MRI technology can be replaced or complemented by EEG is promising. Recent studies suggest the feasibility of reconstructing visual stimuli from brain activity using EEG \cite{benchetrit2023brain}, which offers a more accessible, and cost-effective alternative to MRI. 

Finally, in our experiments, the subject clearly expressed that the imagined images were getting clearer and more precise as he repeatedly attempted to imagine the same image. This opens a potentially promising avenue for "training" one's own mental imagery, which could be applied to subjects presenting various degrees of aphantasia \cite{zeman2015lives}, i.e. the degree of their inability to create mental imagery. A set of experiments directed toward this question is thus a potential future work.

\section{Conclusion}

In our work, we introduced and applied two novel data collection protocols: weak and strong imagination. These protocols were designed to study visual imagery and create training datasets for fMRI-to-Images models. We implemented them during a 6-hour scanning session to create a comprehensive dataset. This dataset then served as the basis for training our fMRI-to-image model. Initially, the model was trained using data from the weak imagination protocol, which involved reconstructing images that the subject had previously seen. Subsequently, the same model, now frozen, was applied to data from the strong imagination protocol, where the subject was asked to imagine new images based on a written instruction. Our study demonstrates that the model successfully distinguishes between broad categories such as portraits and landscapes, effectively grasping the category of the content imagined by the subject. Accurately capturing the detailed contents of these imagined images was partly successful, but proved to be more challenging. This research provides a promising path towards the generation of visual representations directly from human thought.


\bibliography{references}
\bibliographystyle{plain}

\appendix
\onecolumn
\section{Uncurated samples for weak imagination}

\begin{figure*}[ht!]
    \centering
    \includegraphics[scale=0.21]{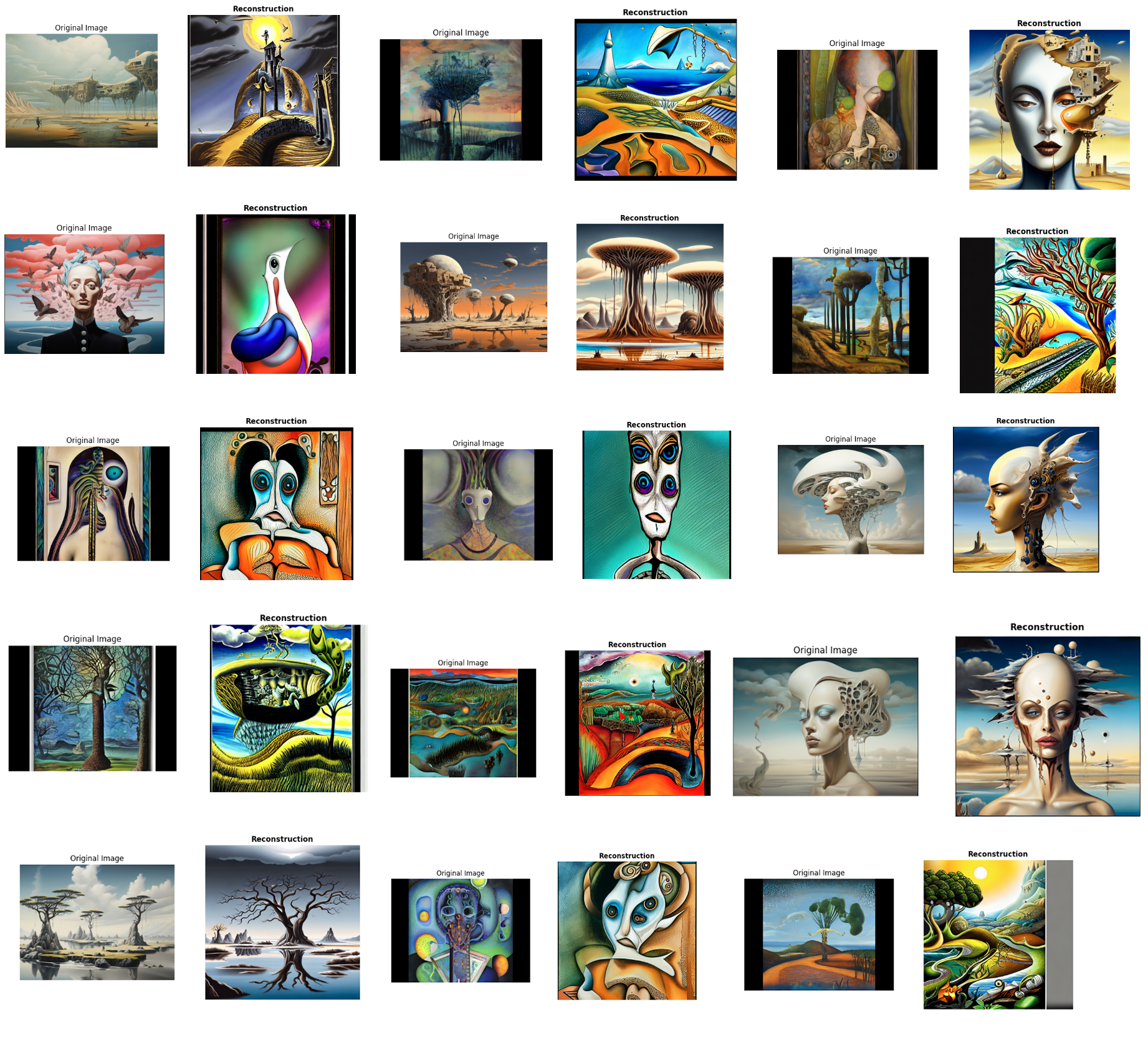}
    \caption{Uncurated results for weak imagination on the validation dataset. In each example, we present an image from the validation set of our surrealism image dataset, which is recollected from memory by the subject in the fMRI scanner, and the reconstructed image from the fMRI-to-Image model based on the brain data associated to the recollection of the original image.}
    \label{fig:uncurated-weak}
\end{figure*}

In Fig.\ref{fig:uncurated-weak}, we present uncurated results from the validation dataset.

\section{Experimental details}

All our models are trained on a single A100 GPU for 500 epochs with a batch size of 16, following \cite{Scotti2023ReconstructingMindsEye} and our constraint of having bigger MLPs in the BrainNetwork class ($h$ was reduced to 2048). Trainings took anywhere from 15 to 25 hours depending on the size of the mask. The optimizers and training routine remained unchanged.

\end{document}